\documentclass[9pt,twocolumn,twoside]{opticajnl}
\journal{opticajournal} % use for journal or Optica Open submissions

\setboolean{shortarticle}{true}
\usepackage{lineno}
\usepackage{verbatim}
\title{Demonstration of a Squeezed Light Source on Thin-Film Lithium Niobate with Modal Phase Matching}

\author[1,$\dagger$]{Tummas Napoleon Arge}
\author[2,3,4,$\dagger$]{Seongmin Jo}
\author[1]{Huy Quang Nguyen}
\author[2,3,*]{Francesco Lenzini}\author[2,3]{Emma Lomonte} 
\author[1]{Jens Arnbak Holbøll Nielsen}
\author[1]{Renato R. Domeneguetti}
\author[1]{Jonas Schou Neergaard-Nielsen}
\author[2,3,4]{Wolfram Pernice}
\author[1,**]{Tobias Gehring}
\author[1,***]{Ulrik Lund Andersen}

\affil[1]{Center for Macroscopic Quantum States bigQ, Department of Physics, Technical University of Denmark, Fysikvej 307, DK-2800 Kgs. Lyngby, Denmark}
\affil[2]{CeNTech-Center for Nanotechnology, Heisenbergstraße 11, 48149, Münster, Germany}
\affil[3]{SoN-Center for Soft Nanoscience, Busso-Peus-Straße 10, 48149, Münster, Germany}
\affil[4]{Kirchoff-Institute of Physics, Heidelberg University, Heidelberg, Germany}
\affil[$\dagger$]{Equal contribution}
\affil[*]{\url{lenzini@uni-muenster.de}}
\affil[**]{\url{tobias.gehring@fysik.dtu.dk}} %% email address is required; see note below about the corresponding author designation

\affil[***]{\url{ulrik.andersen@fysik.dtu.dk}}

\begin{abstract} 
    Squeezed states are essential for continuous variable (CV) quantum information processing, with wide-ranging applications in computing, sensing and communications. Integrated photonic circuits provide a scalable, convenient platform for building large CV circuits. Thin-film Lithium Niobate (TFLN) is particularly promising due to its low propagation loss, efficient parametric down conversion, and fast electro-optical modulation.    
    In this work, we demonstrate a squeezed light source on an integrated TFLN platform, achieving a measured shot noise reduction of 0.46 dB using modal phase matching and grating couplers with an efficiency of up to -2.2 dB. 
    The achieved squeezing is comparable to what has been observed using more complex circuitry based on periodic poling. 
    The simpler design allows for compact, efficient and reproducible sources of squeezed light. 
\end{abstract}

\begin{document}

\maketitle

%%%%%%%%%%%%%%%%%%%%%%%%%%  body  %%%%%%%%%%%%%%%%%%%%%%%%%%
\section{Introduction}
Squeezed states of light are a versatile and indispensable resource for continuous-variable (CV) quantum optical information processing with applications in quantum computing~\cite{Weedbrook2012}, quantum communication~\cite{Pirandola2020} and quantum sensing~\cite{Pirandola2018}. Practical implementations of these protocols require a reliable and scalable source of squeezed light.  
%computing\cite{Asavanant2019,Larsen2019,Madsen2022}, quantum key distribution\cite{Madsen2012, CV_QKD_DI,Li_2023}, and quantum sensing\cite{Yonezawa_2012,Slussarenko_2017,Guo2019}. %variatonal quantum phase estimators\cite{nielsen2023variational}, 
%in addition, squeezed states form the back bone for generating the holy grail in CV quantum computing, Gottesman Knill Preskill states\cite{Konno2024,Takase_2023,feedforward}. 

Quantum optical systems using free space optics on optical benches occupy significant space and suffer from the inherent instability of individual components which hinders the development of large and complex systems. 
Integrated photonics on the other hand offers scalability in the number of components and inherent phase stability through accurate control of the optical path length~\cite{politi2009}. In addition, the high modal confinement of photonic waveguides enables large optical nonlinearities and a strong electro-optic response~\cite{Wang_2019,Loncar_review}.  

Squeezed light sources have been developed on several integrated photonic platforms. 
Using periodically poled titanium-indiffused lithium niobate waveguides up to -3.4 dB of squeezing has been measured~\cite{Domeneguetti:23,Stefszky:23}. However, the large bending radii of the waveguides~\cite{Lenzini2018} limit circuit depth, making this architecture unsuitable for practical applications of CV protocols. Ridge PPLN waveguides bonded on a LiTaO3 substrate fabricated with a mechanical structuring method, have enabled the measurement of squeezing levels up to -8 dB~\cite{Kashiwazaki2021,Kashiwazaki2023}. However, the relatively weak confinement of these waveguides, as well as the employed fabrication method, prevent the realization of highly dense integrated photonic circuits.

Silicon nitride is another promising platform due to its low propagation loss, high index contrast allowing for complex circuitry and CMOS compatibility~\cite{Rahim_2017}. Using microring resonators, up to -1.65 dB of squeezed light has been measured~\cite{Zhang2021}. However, Silicon nitride is limited by the weaker $\chi^{(3)}$ non-linearity, slow modulation speed and parasitic non-linear effects which hinder the performance at high powers needed to generate highly squeezed states~\cite{Rahim_2017}.

Thin film lithium niobate is an appealing platform for integrated CV quantum photonics due to its intrinsic low propagation loss, large electro-optic coefficient, which enables fast electro-optical modulators~\cite{Loncar_review}, and the high $\chi^{(2)}$ nonlinearity, which allows for efficient generation of squeezed states. Using periodic poling, ultra efficient wavelength conversion has been achieved~\cite{Wang2018}. Squeezed states have been generated using both single-pass configuration~\cite{Stokowski2023,Chen:22,Nehra2022} and a cavity ~\cite{park2024single}.
A reduction of the shot noise of up to -0.5 dB has been measured, without correcting for or accounting for any loss induced by the measurement equipment \cite{Nehra2022}. However, periodic poling requires an advanced fabrication process that is not yet mastered by many fabs. It demands a large surface area on the chip and is sensitive to fabrication imperfections, which reduces reproducibility. 

As an alternative, the geometric dispersion of optical transverse modes can be engineered to achieve phase matching between far detuned modes~\cite{Chen:18}. This approach has been utilised to demonstrate second harmonic generation~\cite{Hong2018} and sum frequency generation~\cite{Luo2019}.

In this work we demonstrate a compact squeezed light source on thin film lithium niobate without using periodic poling. Instead, we employ modal phasematching to generate 1550 nm squeezed light in the TE0 fundamental mode while pumping from the higher ordered TM2 775 nm mode.  We measured -0.46 dB of squeezing collected into a fiber, close to the highest reported number with periodic poling. This result is obtained using a compact, all passive device, that does not require any complex circuitry, utilizing  grating couplers with an efficiency of up to -2.2 dB. Being limited by optical fiber coupling loss and the photorefractive effect, we expect that future improvements could lead to the observation of up to -3.6 dB squeezing.

\section{Squeezed Light Source} 
\begin{comment}
    
\begin{figure}[ht]
  \centering
  \includegraphics[width=\columnwidth]{Figures/two columns/scheme.pdf}
    \caption{a) Scheme to produce squeezed light using modal phase matching on Z-cut thin film lithium niobate. The TM0 input 775 nm mode excites a TM2 mode inside the resonator that is converted to to TE0 1550 nm squeezed vacuum, which is coupled out to from the cavity. b) The waveguide modes that are phasematched, from Meep. c) Phasematching is achieved at a top width of around 1 µm, simulation from Lumerical.}\label{fig:scheme}
    %}
\end{figure}
\end{comment}

\begin{figure}[ht]
  \centering
  \includegraphics[width=\columnwidth]{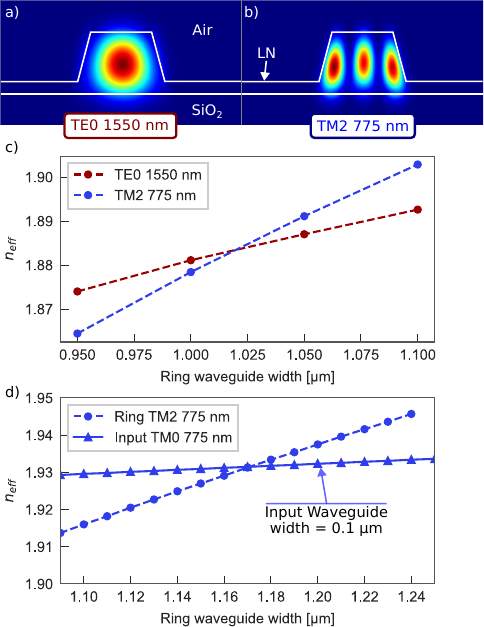}
    \caption{Modal power profile of the a) TE0 1550 nm  mode and b) TM2 775 nm mode. c) Phasematching is achieved at a top width of around 1 µm. d) The TM2 775 nm mode is excited from the TM0 775 nm mode in the 0.1 µm wide input waveguide. All simulations are made using  Lumerical. }\label{fig:simulations}
    %}
\end{figure}

To achieve phase matching between the 775 nm pump light and the 1550 nm signal light, we leverage the birefringence and geometric dispersion to counteract the intrinsic dispersion between the two wavelengths. By using a Z-cut LN thin film, where the extraordinary refractive index lies in the vertical plane, we achieve an isotropic refractive index in the device plane. To minimize propagation loss, we design for phasematching between the fundamental quasi-transverse-electric mode (TE0) at 1550 nm, Fig.~\ref{fig:simulations}~a, and the high-order quasi-transverse-magnetic mode (TM2) at 775 nm, Fig.~\ref{fig:simulations}~b, allowing for a greater waveguide width. This design employs the weaker $d_{31}$ nonlinear term. Phase matching is achieved between these modes at around a 1 µm waveguide width, as shown in Fig.~\ref{fig:simulations} c, where we plot the effective refractive indices for the 1550\,nm TE0 and the 775\,nm TM2 mode versus the waveguide width. Phase-matching occurs when both modes have the same effective refractive index. 

To enhance the non-linear efficiency, we employ a ring resonator cavity that is resonant at both wavelengths. The radius of the cavity is set to 70 µm, balancing low propagation loss and high finesse. %The device dimensions are 580 µm x 520 µm (W x H).
To augment the coupling strength at a wavelength of 775 nm, a pulley coupler is employed to convert the mode from TM0 to TM2. This mode conversion is achieved by matching the effective refractive indices between the waveguide and cavity mode \cite{Lu2019P}. %This mode excites a higher ordered TM2 mode in a resonator using a pulley coupler. By optimising the width in the bottom bend, the TM2 mode can be made to phase match the fundamental TE0 1550 nm signal mode. Modal phasematching between these modes is achieved with a top width of 1 µm. Both wavelengths are resonant in the resonator.

\begin{figure}[ht]
  \centering
  \includegraphics{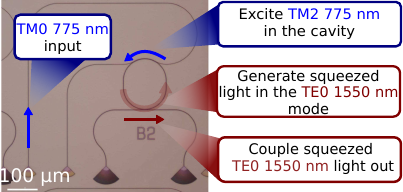}
    \caption{Device used to generate squeezed light using modal phasematching }\label{fig:scheme}
    %}
\end{figure}

The reproducible minimum waveguide thickness during the fabrication process was 0.1 µm, and the cavity thickness needed to be 1.16 µm to achieve effective refractive index matching with the TM2 mode, as shown in Fig.~\ref{fig:simulations} d. Given that the waveguide thickness for modal phase matching was 1.02 µm, the cavity was designed with a thickness of 1.16 µm at the top for mode conversion and 1.02 µm at the bottom for phase matching. A linear waveguide taper of 25 µm was used between these sections. The final device, shown in Fig.~\ref{fig:scheme}, has dimensions of 580 µm x 520 µm (W x H).

The fabrication process started with a $15\times15  \text{mm}^2$ diced chip from a NANOLN wafer with a 500 nm thick Z-cut lithium niobate film bonded to an 4.7 µm thick silicon dioxide insulating layer thermally grown on silicon. After spin-coating with a negative-tone ArN 7520 resist, the chip underwent electron beam lithography patterning. Subsequent development was performed using an MF-319 solution, after which patterns were transferred onto the LN thin films through a physical sputtering process using argon plasma in an Oxford100 ICP-RIE etching tool with an etch depth of 0.40 µm. To address the issue of material redeposition on the waveguide surfaces after etching, wet etching (RCA-1 cleaning) was performed. Finally, to reduce propagation loss, the sample was annealed at $400^{\circ}$C.

For light coupling onto the chip, negative angle grating couplers were used. These grating couplers diffract incident light from an optical fiber into an on-chip waveguide and vice versa. By optimising the structural periodicity of the grating, we achieved phase-matching between the incoming light in the fiber mode and the waveguide mode. The efficiency of the grating coupler was highly wavelength-dependent.

%The structural periodicity of the grating is critical as it dictates the coupling angle and efficiency through fulfillment of the phase matching condition between the incoming light and the waveguide mode. The efficiency of grating couplers is notably wavelength-dependent, necessitating precise adjustment of design parameters to optimize their performance at designated wavelengths.

\begin{figure}[htbp]
    \centering
    \includegraphics[width=\columnwidth]{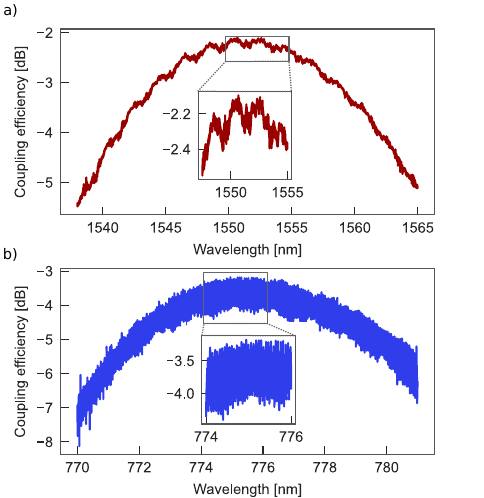}
    \caption{a) Measured coupling efficiency for the signal. It exhibits an efficiency of -2.2 dB at a wavelength of 1550 nm. b) Measured coupling efficiency for the pump. At the optimal height, an efficiency of -3.67 dB was measured at a wavelength of 775 nm. During sweeping of the wavelength, instability in the laser power resulted in noise. }
    \label{fig:transmission}
\end{figure}

%The scheme for generating squeezed light is seen in figure \ref{fig:scheme} a) and b). Using grating couplers a TM0 775 nm pump mode is coupled to the chip. Using a width of 0.1 µm for the input waveguide and 1.016 µm for the top half of the cavity waveguide, the TM2 mode is excited from the TM0 input pump mode, figure \ref{fig:scheme} c). The bottom half of the cavity has a top width of 1020 nm to achieve phasematching between the TM2 775 nm pump mode and the fundamental 1550 nm signal mode, figure \ref{fig:scheme} d). To optimise the measured phasematching, we sweep the waveguide width on the chip. We couple the generated squeezed light out by using grating couplers. 

A measurement of the coupling efficiency is shown in Fig.~\ref{fig:transmission}. For the 1550 nm light, a loss of 2.2 dB per facet was measured, while the pump light at 775 nm exhibited a loss of 3.67 dB per facet. However, when using a fiber array to simultaneously couple both wavelengths, the efficiency for the 775nm pump light was measured to -5.8 dB at the maximum efficiency of the 1550 nm signal light. This discrepancy is due to a difference in focusing height. 

We designed the cavity to be double resonant. To maximise the squeezing, the cavity was overcoupled for the signal light at 1550 nm, allowing as much squeezed light to exit the cavity as possible. The Q-factor of this cavity was $1.5\cdot10^5$. To achieve maximal efficiency of the non-linear process, we aimed for a critically coupled cavity for the 775 nm pump light, resulting in a Q-factor of $7.1\cdot10^4$. 

%The devices used to generate squeezing can be seen in figure \ref{fig:scheme}. Using grating couplers a TM0 775 nm pump mode is coupled to the chip. This mode is excites a higher ordered TM2 mode in a ring resonator using a pulley coupler. By optimising the width in the bottom bend, the TM2 mode can be made to phase match the fundamental TE0 1550 nm signal mode. Modal phasematching between these modes is achieved with a top width of 1 µm. 

%for the pump wavelength, due to the difference in focusing height with the signal wavelength, a coupling efficiency of 26 \% was observed at the maximum efficiency of the signal wavelength. % not sure if we should change this to explain 

\begin{figure}[htbp]
    \centering
    \includegraphics[width=\columnwidth]{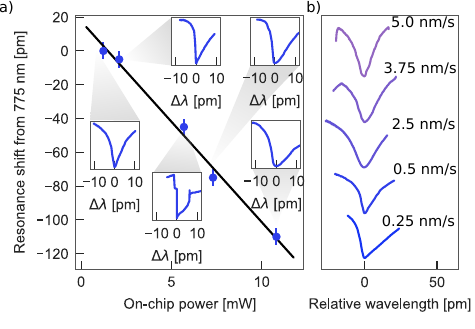}
    \caption{a) Effect of power on the pump resonance. The photorefractive effect transforms the resonance from a Lorentzian to a shark fin in addition to blue-shifting the light. The data is taken at 0.5 nm/s. b) increasing the scan speed decreases the photorefractive effect. At high scan speeds the Lorentzian lineshape is recovered. The resonances are normalized and y-shifted. } %
    \label{fig:resonances_775}
\end{figure}

% move b) for more space, remove y label, caption

Devices on a Lithium niobate integrated platform experience a strong photorefractive effect~\cite{Photorefractive}, where charge migration due to optical illumination causes an electro-optic effect\cite{Weis1985,HALL198577}. Due to the weak interaction of the TM2 pump mode and the TE0 signal mode, a strong pump field is required, which in turn induces a significant photorefractive effect, as explored in Fig.~\ref{fig:resonances_775}. This effect results in a linear blue shift of the resonance, as seen in Fig. \ref{fig:resonances_775}a, with a slope of -17.4 nm/mW at a scan speed of 0.5 nm/s. Instead of a Lorentzian resonance shape, the photorefractive effect produces a characteristic "shark fin" transmission spectrum, which becomes more pronounced with increasing power. Up to 5 mW, this shark fin effect becomes more distinct, but beyond 5 mW, the resonance becomes distorted and wider.  

The resonance stretching due to the photorefractive effect is inversely proportional to the scan-speed of the laser wavelength~\cite{Photorefractive}. In Fig.~\ref{fig:resonances_775}b, we show the measured transmission spectrum of the pump when scanning the wavelength around the resonance with a pump power of 0.9 mW on-chip. As seen in the figure, the characteristic Lorentzian resonance shape is recovered at high scan speeds, while the shark fin resonance is clearly visible for low scan speeds, even at low powers, which is consistent with the findings of Shen et. al. \cite{Photorefractive}. The effect on the 1550 nm signal mode is minimal and therefore not shown in the figure. 

\section{Parametric Gain}
\begin{figure}[htbp]
    \centering
    \includegraphics[width=\columnwidth]{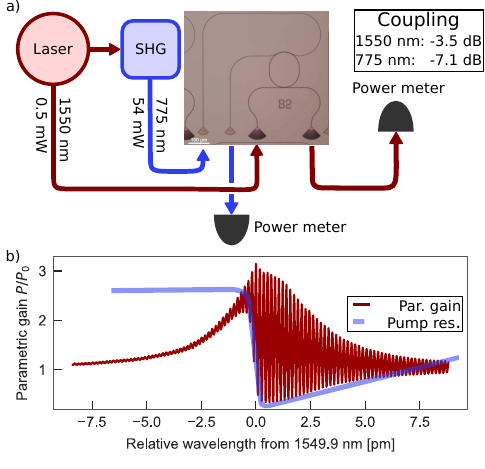}
    \caption{a) Circuit for measuring parametric gain. b) Observed parametric gain, the ripples are due to the pump light oscillating between being in and out of phase with the seed light. The sharkfin pump envelope (blue) distorts the gain. The sweep speed is 1 nm/s. }
    \label{fig:gainPlot}
\end{figure}
%\subsection{Classical characterisation}
To characterize the squeezed light source, we first measured its parameteric gain. The experimental setup for this measurement is shown in Fig.~\ref{fig:gainPlot}a. An NKT Boostik E15 laser supplied 1550 nm light, part of which was up-converted to 775 nm light using an NKT Harmonik module. The chip was mounted on an aluminum block with a Peltier element for temperature control. The light was monitored with a Thorlabs PDA30B2 photodetector for the 1550 nm and a Thorlabs PDA36A2e for the 775 nm path. By sweeping the laser frequency and laser power and optimising the temperature of the chip, double resonance and phase matching was achieved. 

A typical measurement of the parametric gain is shown in Fig.~\ref{fig:gainPlot}b which was obtained by amplifying a weak 1550 nm seed beam with a strong 775 nm pump field. The gain was estimated by comparing the transmitted 1550 nm light with and without the pump. 
%The gain constant is modelled as % need somehow to put in calculation? 
%\begin{gather}
%    g^2 = \frac{1}{\left(1-\frac{P_p}{P_{p,th}}\right)^2} = \frac{P_s}{P_{s,0}}
%\end{gather}
%where $P_p$ is the input pump power and $P_{p,th}$ is the OPO threshold power. $P_s$ is the measured 1550 nm power and the $P_{s,0}$ is the 1550 nm power without any input 775 nm light. 
When scanning the wavelength over a resonance, the pump light interacts in and out of phase with the seed light, leading to the ripples seen in the figure. % By optimising the pump power, center wavelength and temperature of the chip the gain constant is 3, figure \ref{fig:gainPlot} c). %From this the OPO threshold can be estimated to be $\infty$ W. 
The photorefractive effect, which causes the shark fin envelope described above and shown in Fig.~\ref{fig:resonances_775}, distorts the parametric gain envelope because it shifts the center resonance frequency for the pump light, but not for the signal light. Consequently, the double resonance condition is only fulfilled for specific optical powers. 

With 10 mW of on-chip power, the centre of the 1550 nm resonance exhibited an amplification factor of 3.15 and a deamplification factor of 0.5, taken as the highest and lowest value of the ripples. This corresponds to a threshold power of the OPO of around 50 mW of on-chip power.

\section{Squeezing}
\begin{figure*}[htbp]
    \centering
    \includegraphics[width=\textwidth]{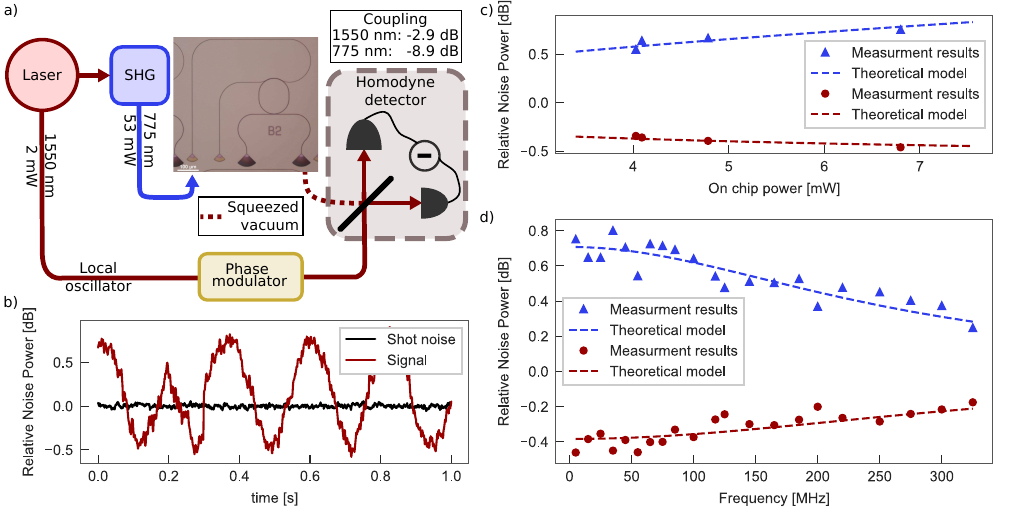}
    \caption{a) Circuit for generating squeezed light. The photo diodes of the homodyne detector have a quantum efficiency of $\eta=0.85$. b) Squeezing and anti-squeezing at a sideband frequency of 5 MHz with an on-chip power of 6.9 mW. The resolution bandwidth is 1 MHz and the video bandwidth is 100 Hz. c) Measured noise power relative to shot noise versus pump power (775 nm) at a sideband frequency of 5 MHz. d) Measured noise power relative to shot noise versus sideband frequency for an on-chip power of 6.9 mW at 775 nm.  }
    \label{fig:sqzPlot}
\end{figure*}

\begin{comment}
\begin{figure}[htbp]
    \centering
    \includegraphics[width=\columnwidth]{Figures/two columns/squeezing_trace.pdf}
    \caption{a) Circuit for generating squeezed light. The photo diodes of the homodyne detector have a quantum efficiency of $\eta=0.85$. b) Squeezing and anti-squeezing at a sideband frequency of 5 MHz with an on-chip power of 6.9 mW. The resolution bandwidth is 1 MHz and the video bandwidth is 100 Hz. }
    \label{fig:sqzPlot}
\end{figure}
\end{comment}

We then used the experimental setup shown in Fig.~\ref{fig:sqzPlot}a to measure the squeezing generated by the thin film lithium niobate resonator. The 1550 nm light was split in two parts: the majority was used to generate a strong 775 nm beam, while the remainder was used as a local oscillator (LO) for homodyne detection. The phase of the LO light was scanned at 0.5 Hz with respect to the squeezed beam. The coupling efficiency was reduced from the -2.2 dB to -2.9 dB compared to Fig.~\ref{fig:transmission} due to damage to the devices. 

Figure \ref{fig:sqzPlot}b shows a typical zero-span measurement of the noise power at 5 MHz from the homodyne detector. With 6.9 mW of pump power on the chip, the measured squeezing and anti-squeezing levels are -0.46 dB and 0.75 dB, respectively. The shot noise level was determined by blocking the output from the lithium niobate chip. From the squeezing and anti-squeezing measurement results, we estimated the total efficiency of the setup to be approximately $\eta \approx 0.24$.

Next, we tuned the 775\,nm pump power, and the results of squeezing measurements are shown in Fig.~\ref{fig:sqzPlot}c. Due to the photorefractive effect on the 775\,nm resonance, double resonance had to be individually achieved for each data point. While the wavelength was fixed to achieve resonance for the 1550\,nm light, the resonance for 775\,nm was obtained by adjusting the temperature of the circuit and finely tuning the pump power. This provided only a small region of pump power where double resonance could be achieved; outside of the region no squeezing was measured, explaining the sparse data points in the figure. In addition, the stability of the coupling of pump light on chip was crucial, as variations in the pump light shifted the pump resonance away from the signal resonance due to the photorefractive effect. Each data point in the figure represents an individual scanned LO measurement, where the value was determined by taking the mean of 30 data points around the minimum (maximum) of squeezing (anti-squeezing) of the signal. 

At a lower on-chip pump power of 4.0 mW, compared to the previously discussed 6.9\,mW, the squeezing and anti-squeezing levels decreased to -0.34 dB and 0.55 dB, respectively. We fitted a theoretical model to the measurement data. The noise power of the squeezing, $S_{-}$, and anti-squeezing, $S_{+}$, are given by $S_\pm = 1\pm\eta\frac{4\sqrt{\frac{P_p}{P_p^{th}}}}{\left(1\mp\sqrt{\frac{P_p}{P_p^{th}}}\right)^2+\left(\frac{f}{\bar{f}_s} \right)^2}$, where the total efficiency is $\eta = \eta_{QE}\mathcal{V}^2\eta_{opt}\eta_{esc}$, $P_p$ is the pump power, $P_p^{th}$ is the oscillation threshold of the oscillator, $\omega$ is the side-band frequency, and $\Bar{f}_s$ is the HWHM of the cavity bandwidth of the signal in natural units. From the fit we obtained a total efficiency $\eta=0.20$ and an on-chip pump threshold of $P_{th}\approx 200$\,mW, higher than the 50 mW estimated from the parametric gain. This higher number can be attributed to the difficulty achieving double resonance. Clearly, the pump powers used in our experiment to measure squeezing are well below the estimated threshold power.

%\begin{gather}
%    S_\pm = 1\pm\eta\frac{4\sqrt{\frac{P_p}{P_p^{th}}}}{\left(1\mp\sqrt{\frac{P_p}{P_p^{th}}}\right)^2+\left(\frac{\omega}{\Gamma_s} \right)^2},\qquad \eta = \eta_{QE}\mathcal{V}^2\eta_{opt}\eta_{esc}\ ,
%\end{gather}

%By decreasing the input pump power, the squeezing and antisqueezing levels are decreased, Fig.~\ref{fig:sqzPlot}c. The decrease is small as the pump power is far away from the oscillation threshold of the resonator. At 4.0 mW of on-chip power the squeezing and anti-squeezing levels are decreased -0.34 dB and 0.55 dB, respectively. In contrast to the measurement of parametric gain, for this measurement the laser wavelength must be fixed to the resonances of the pump and signal. Double resonance is achieved by changing the temperature and the pump power. This gives a small region of pump power where double resonance is achieved, outside of which no squeezing can be measured. In addition, the coupling of pump light on chip must be kept stable, as a loss of pump light moves the pump resonance away from the signal resonance due to the photorefractive effect. Each data point in the figure stems from an individual measurement, where the value was determined by taking the mean of 30 data points around the minimum (maximum) of squeezing (anti-squeezing) of the signal when scanning the local oscillator phase. The variation of the squeezing and anti-squeezing values can be explained by the different coupling efficiencies of the individual measurements.

Finally, we measured squeezing and anti-squeezing as a function of frequency at a pump power of 6.9 mW. The results are shown in Fig.~\ref{fig:sqzPlot}d. Since the LO phase was not locked, we performed individual zero-span measurements for each data point, similar to the measurements versus pump power. The relatively large variation of the squeezing and anti-squeezing values can be attributed to different coupling efficiencies during individual measurements. As expected, a clear reduction of squeezing and anti-squeezing with increasing sideband frequency is observed, due to the finite bandwidth of the cavity. In comparison to the previous measurements at a 5 MHz sideband frequency, the squeezing and anti-squeezing levels at 325 MHz decreased to -0.18 dB and 0.25 dB, respectively. The fit of the theoretical model to the measurement data is shown in the figure with the following parameters: $P_p / P_p^{th}=0.02$, total efficiency $\eta=0.23$, and cavity bandwidth $\bar{f}_s=310$ MHz. The total efficiency $\eta$ is determined by the quantum efficiency of the detectors $\eta_{QE}=0.85$, the visibility of the homodyne detector $\mathcal{V}^2\approx0.98$, and the optical efficiency $\eta_{opt}=0.45$. Lastly, the escape efficiency is $\eta_{esc}\approx 0.55$

\section{Discussion and outlook}
In summary, we have developed a compact source of squeezed light by employing modal phasematching on a thin film lithium niobate platform. By optimising the geometry of the waveguide circuit, we enabled the TM2 pump mode to propagate with the same effective index as the TE0 signal mode. This configuration provided a gain coefficient of 3.16, allowing us to measure light squeezed -0.46 dB below shot noise. This result matches the highest measured squeezing generated in thin film lithium niobate \cite{Stokowski2023,park2024single}, achieved without the use of periodic poling and by employing an all passive device. 

The performance of our devices was limited by three factors: 1) coupling loss, 2) the photorefractive effect of the pump light, and 3) the difficulty of simultaneously attaining resonance for both the pump and signal wavelengths. 

Addressing the coupling loss, 1), inherent to thin film lithium niobate circuits, is a significant engineering challenge. Grating couplers with a negative angle, as used here, have the potential to reach coupling efficiencies close to -0.1 dB if complemented with metal back-reflectors \cite{Lomonte_2024}. We estimate that the generated squeezing on-chip was -1.5 dB, significantly higher than the measured off-chip value. The photorefractive effect, 2), is intrinsic to lithium niobate. To mitigate this, the device's operating temperature can be increased \cite{Domeneguetti:23}, or MgO doping can be employed \cite{Fonseca_Campos_2007}. Another option is to increase the length of the cavity, thereby decreasing the resonance power buildup of the pump light, thereby decreasing the photorefractive effect.
For moderate increases in length, the extended phase-matched region will compensate for the decrease in resonance, enhancing the nonlinear efficiency. A decrease in photorefractive effect, 2), will give a more gentle slope for the blue-shift of the resonance. Thus it can be used to tune the device to double resonance, 3), making it possible to achieve oscillation threshold power closer to the 50 mW estimated from the gain measurement. 

%Paradoxically this will reduce the tunability of the devices, as the photorefractive effect also moves the pump resonance thereby increasing 3). 

The limited tunability could also be addressed by adding a local heater or an electro-optic modulator, mitigating the loss of tunability by increasing the length of the resonator. 
We estimate that if we were able to pump with a power close to the threshold value, -1.2 dB of squeezing could be measured off-chip with the current grating couplers and -3.6 dB on chip. %This may be possible by improving 2) and 3). %By addressing 3) it may also be possible to optimise the temperature of the peltier to maximise the phasematching and to achieve double resonance. This would increase the non-linear generation. 

Developing high quality squeezed light sources is a challenging task, but worth the effort given the high prospects in terms of applications of on-chip squeezed light sources. This work is a step towards this goal.

\begin{backmatter}
\bmsection{Funding}
The research was funded by the Danish National Research Foundation, Center for Macroscopic Quantum States (bigQ, DNRF142), the Innovation Fund Denmark (PhotoQ project, no. 1063-00046A), and from the EU projects CLUSTEC (grant agreement no. 101080173) and EPIQUE (grant agreement no. 101135288). 
%\textcolor{red}{Maybe you need to add some more funding agencies here Seongmin, idk?}

\bmsection{Acknowledgments}
We thank Daniel A. Requena and Anders J. E. Bjerrum for fruitful discussions.

\bmsection{Disclosures}
The authors declare no conflicts of interest.

\bmsection{Data Availability Statement}
All relevant data is included in the paper. The data for the squeezing measurements can be found here: \url{https://figshare.com/s/cbf2154fa18e9201faab} %need to change for later as it is a private link. It needs a proper DOI
\end{backmatter}

\bibliography{bibliography}

\begin{thebibliography}{10}
\newcommand{\enquote}[1]{``#1''}

\bibitem{Weedbrook2012}
C.~Weedbrook, S.~Pirandola, R.~García-Patrón, \emph{et~al.}, {\protect\JournalTitle{Reviews of Modern Physics}} \textbf{84}, 621–669 (2012).

\bibitem{Pirandola2020}
S.~Pirandola, U.~L. Andersen, L.~Banchi, \emph{et~al.}, {\protect\JournalTitle{Advances in Optics and Photonics}} \textbf{12}, 1012 (2020).

\bibitem{Pirandola2018}
S.~Pirandola, B.~R. Bardhan, T.~Gehring, \emph{et~al.}, {\protect\JournalTitle{Nature Photonics}} \textbf{12}, 724–733 (2018).

\bibitem{politi2009}
A.~Politi, J.~C. Matthews, M.~G. Thompson, and J.~L. O'Brien, {\protect\JournalTitle{IEEE Journal of Selected Topics in Quantum Electronics}} \textbf{15}, 1673 (2009).

\bibitem{Wang_2019}
J.~Wang, F.~Sciarrino, A.~Laing, and M.~G. Thompson, {\protect\JournalTitle{Nature Photonics}} \textbf{14}, 273–284 (2019).

\bibitem{Loncar_review}
D.~Zhu, L.~Shao, M.~Yu, \emph{et~al.}, {\protect\JournalTitle{Advances in Optics and Photonics}} \textbf{13}, 242 (2021).

\bibitem{Domeneguetti:23}
R.~Domeneguetti, M.~Stefszky, H.~Herrmann, \emph{et~al.}, {\protect\JournalTitle{Opt. Lett.}} \textbf{48}, 2999 (2023).

\bibitem{Stefszky:23}
M.~Stefszky, F.~vom Bruch, M.~Santandrea, \emph{et~al.}, {\protect\JournalTitle{Opt. Express}} \textbf{31}, 34903 (2023).

\bibitem{Lenzini2018}
F.~Lenzini, J.~Janousek, O.~Thearle, \emph{et~al.}, {\protect\JournalTitle{Science Advances}} \textbf{4} (2018).

\bibitem{Kashiwazaki2021}
T.~Kashiwazaki, T.~Yamashima, N.~Takanashi, \emph{et~al.}, {\protect\JournalTitle{Applied Physics Letters}} \textbf{119} (2021).

\bibitem{Kashiwazaki2023}
T.~Kashiwazaki, T.~Yamashima, K.~Enbutsu, \emph{et~al.}, {\protect\JournalTitle{Applied Physics Letters}} \textbf{122} (2023).

\bibitem{Rahim_2017}
A.~Rahim, E.~Ryckeboer, A.~Z. Subramanian, \emph{et~al.}, {\protect\JournalTitle{Journal of Lightwave Technology}} \textbf{35}, 639–649 (2017).

\bibitem{Zhang2021}
Y.~Zhang, M.~Menotti, K.~Tan, \emph{et~al.}, {\protect\JournalTitle{Nature Communications}} \textbf{12} (2021).

\bibitem{Wang2018}
C.~Wang, C.~Langrock, A.~Marandi, \emph{et~al.}, {\protect\JournalTitle{Optica}} \textbf{5}, 1438 (2018).

\bibitem{Stokowski2023}
H.~S. Stokowski, T.~P. McKenna, T.~Park, \emph{et~al.}, {\protect\JournalTitle{Nature Communications}} \textbf{14} (2023).

\bibitem{Chen:22}
P.-K. Chen, I.~Briggs, S.~Hou, and L.~Fan, {\protect\JournalTitle{Opt. Lett.}} \textbf{47}, 1506 (2022).

\bibitem{Nehra2022}
R.~Nehra, R.~Sekine, L.~Ledezma, \emph{et~al.}, {\protect\JournalTitle{Science}} \textbf{377}, 1333–1337 (2022).

\bibitem{park2024single}
T.~Park, H.~Stokowski, V.~Ansari, \emph{et~al.}, {\protect\JournalTitle{Science Advances}} \textbf{10}, eadl1814 (2024).

\bibitem{Chen:18}
J.-Y. Chen, Y.~M. Sua, H.~Fan, and Y.-P. Huang, {\protect\JournalTitle{OSA Continuum}} \textbf{1}, 229 (2018).

\bibitem{Hong2018}
L.-H. Hong, B.-Q. Chen, C.-Y. Hu, and Z.-Y. Li, {\protect\JournalTitle{Physical Review A}} \textbf{98} (2018).

\bibitem{Luo2019}
R.~Luo, Y.~He, H.~Liang, \emph{et~al.}, {\protect\JournalTitle{Physical Review Applied}} \textbf{11} (2019).

\bibitem{Lu2019P}
J.~Lu, J.~B. Surya, X.~Liu, \emph{et~al.}, {\protect\JournalTitle{arXiv: Optics}}  (2019).

\bibitem{Photorefractive}
Y.~Xu, M.~Shen, J.~Lu, \emph{et~al.}, {\protect\JournalTitle{Opt. Express}} \textbf{29}, 5497 (2021).

\bibitem{Weis1985}
R.~S. Weis and T.~K. Gaylord, {\protect\JournalTitle{Applied Physics A Solids and Surfaces}} \textbf{37}, 191–203 (1985).

\bibitem{HALL198577}
T.~Hall, R.~Jaura, L.~Connors, and P.~Foote, {\protect\JournalTitle{Progress in Quantum Electronics}} \textbf{10}, 77 (1985).

\bibitem{Lomonte_2024}
E.~Lomonte, M.~Stappers, L.~Krämer, \emph{et~al.}, {\protect\JournalTitle{Scientific Reports}} \textbf{14} (2024).

\bibitem{Fonseca_Campos_2007}
J.~Fonseca-Campos, Y.~Wang, W.~Liang, \emph{et~al.}, \enquote{Comparison of photorefractive effects in undoped and mgo-doped ppln,} in \emph{Photonics North 2007,}  (2007).

\end{thebibliography}

\end{document}